\newif\iffigs\figstrue
\figsfalse

\documentstyle[12pt]{article}
\iffigs
  \input epsf
\else
\fi

\batchmode
  \newfont{\footscrfont}{rsfs10}
  \newfont{\footbbbfont}{msbm10}
\errorstopmode

\newif\ifscrf\scrftrue
\ifx\footscrfont\nullfont
  \scrffalse
\fi

\newif\ifamsf\amsftrue
\ifx\footbbbfont\nullfont
  \amsffalse
\fi

\def\ppnumber{\vbox{\baselineskip14pt\hbox{CK-TH-99-001, SUSX-TH-98-014}
\hbox{hep-th/9901150}}}
\def\ppdate{January 1999}
\def\pplogo{\vbox{\kern-\headheight\kern -15pt
\halign{##&##\hfil\cr&{
\ppnumber}\cr\rule{0pt}{2.5ex}&\ppdate\cr}
}}
\makeatletter
\date{}
\def\dedicatory#1{\def\@date{\normalsize\it#1}}
\def\subjclass#1{\def\@thefnmark{}\@footnotetext{1991
    {\it Mathematics Subject Classification.} #1}}
\def\keywords#1{\def\@thefnmark{}\@footnotetext{
    {\it Key words and phrases.} #1}}

\def\ps@firstpage{\ps@empty \def\@oddhead{\hss\pplogo}%
  \let\@evenhead\@oddhead 
}
\def\maketitle{\par
 \begingroup
 \def\thefootnote{\fnsymbol{footnote}}
 \def\@makefnmark{\hbox
 to 0pt{$^{\@thefnmark}$\hss}}
 \if@twocolumn
 \twocolumn[\@maketitle]
 \else \newpage
 \global\@topnum\z@ \@maketitle \fi\thispagestyle{firstpage}\@thanks
 \endgroup
 \setcounter{footnote}{0}
 \let\maketitle\relax
 \let\@maketitle\relax
 \gdef\@thanks{}\gdef\@author{}\gdef\@title{}\let\thanks\relax}

\def\abstract{\if@twocolumn
\section*{Abstract}
\else \small
\begin{center}
{\bf ABSTRACT}
\end{center}
\quotation
\fi}

\def\thebibliography#1{\section*{References\@mkboth
 {REFERENCES}{REFERENCES}}\small\list
 {[\arabic{enumi}]}{\settowidth\labelwidth{[#1]}\leftmargin\labelwidth
 \advance\leftmargin\labelsep
 \usecounter{enumi}}
 \def\newblock{\hskip .11em plus .33em minus .07em}
 \sloppy\clubpenalty4000\widowpenalty4000
 \sfcode`\.=1000\relax}

\newif\iffn\fnfalse

\@ifundefined{reset@font}{\let\reset@font\empty}{} 
\long\def\@footnotetext#1{\insert\footins{\reset@font\footnotesize
    \interlinepenalty\interfootnotelinepenalty
    \splittopskip\footnotesep
    \splitmaxdepth \dp\strutbox \floatingpenalty \@MM
    \hsize\columnwidth \@parboxrestore
   \edef\@currentlabel{\csname p@footnote\endcsname\@thefnmark}\@makefntext
    {\rule{\z@}{\footnotesep}\ignorespaces
      \fntrue#1\fnfalse\strut}}}

\makeatother

\ifamsf
  \newfont{\bigbbbfont}{msbm10 scaled\magstep2}
  \newfont{\bbbfont}{msbm10 scaled\magstep1}  
  \newfont{\smallbbbfont}{msbm8}
  \newfont{\tinybbbfont}{msbm6}
  \newfont{\smallfootbbbfont}{msbm7}
  \newfont{\tinyfootbbbfont}{msbm5}
\fi

\ifscrf
  \newfont{\scrfont}{rsfs10 scaled\magstep1}  
  \newfont{\smallscrfont}{rsfs7}
  \newfont{\tinyscrfont}{rsfs7}
  \newfont{\smallfootscrfont}{rsfs7}
  \newfont{\tinyfootscrfont}{rsfs7}
\fi

\ifamsf
  \newcommand{\Bbb}[1]{\iffn
      \mathchoice{\mbox{\footbbbfont #1}}{\mbox{\footbbbfont #1}}
      {\mbox{\smallfootbbbfont #1}}{\mbox{\tinyfootbbbfont #1}}\else
      \mathchoice{\mbox{\bbbfont #1}}{\mbox{\bbbfont #1}}
      {\mbox{\smallbbbfont #1}}{\mbox{\tinybbbfont #1}}\fi}
\else
  \def\bigbbbfont{\bf}
  \def\Bbb{\bf}
\fi

\ifscrf
  \newcommand{\Scr}[1]{\iffn
    \mathchoice{\mbox{\footscrfont #1}}{\mbox{\footscrfont #1}}
    {\mbox{\smallfootscrfont #1}}{\mbox{\tinyfootscrfont #1}}\else
    \mathchoice{\mbox{\scrfont #1}}{\mbox{\scrfont #1}}
    {\mbox{\smallscrfont #1}}{\mbox{\tinyscrfont #1}}\fi}
\else
  \def\Scr{\cal}
\fi

\def\P{{\Bbb P}}

\def\Z{{\Bbb Z}}

\def\opeq#1{\advance\lineskip#1 \advance\baselineskip#1
        \advance\lineskiplimit#1}

\def\cM{{\Scr M}}

\def\cD{{\Scr D}}

\def\cMc{{\hfuzz=100cm\hbox to 0pt{$\;\overline{\phantom{X}}$}\cM}}
\def\barcD{{\hfuzz=100cm\hbox to 0pt{$\;\overline{\phantom{X}}$}\cD}}

\ifamsf

\else

\bwgin{document}
\fi

\begin{document}
\setcounter{page}0
\title{\LARGE F-Theories On Double Sextics and Effective String 
Theories\\[10mm]}
\author{
Christos Kokorelis\\[0.1cm]
\normalsize Hanover Court, Wellington Road\\
\normalsize BN2 3AZ, Brighton, U.K\\[5mm]
}

{\hfuzz=10cm\maketitle}

\def\Large{\large}
\def\LARGE{\large\bf}

\vskip 1cm

\begin{abstract}
We construct new F-theory vacua in 8-dimensions. They are coming
by projective realizations of F-theory
on K$_3$ surfaces admitting double covers onto $\P^2$, branched 
along a plane sextic curve, the so called double sextics.
The new vacua are associated with singular $K_3$ surfaces.
In this way the stable picture of the heterotic string is mapped
at the triple points of the sextic.
We argue that this formulation incorporates naturally the
$Sp(4,Z)$ invariance that the extrapolating four dimensional
vector multiplet sector of all heterotic vacua may possess.  
In addition, we describe the way that the 4${\cal D}$ g=2 
description of (0,2) moduli dependence of $N=1$ gauge coupling constants
may be connected to Riemann surfaces, with natural
$Sp(4,Z)$ duality invariance. Here we recover a novel way to break
space-time supersymmetry and fix the moduli parameters in the
presence of Wilson lines.
In the context of arithmetic of torsion points on elliptic curves, we
describe in detail, the derivation of the elliptic fibrations in Weierstrass 
form. We also consider the heterotic duals to compactifications of F-theory 
in four dimensions belonging to isomorphic classes of elliptic curves with 
points-cusps of order two. For the latter theories, we calculate the 
${\cal N}=2$ 4D heterotic prepotential f$_{TTT}$ corresponding 
to ${\Gamma_0(2)}_T \times {\Gamma_o(2)}_U$  classical perturbative
duality group and their conjugate modular theories.

\end{abstract}
\vfil\break

\newpage

\section{Introduction and Motivation}

At the current state of the art the five perturbative string theories 
(PST's)
are connected among themselves through the various dualities.
In addition, they are regarded that they originate from compactifications 
of the same underlying higher dimensional theory, M- of F-theory.
In this way, different
choise of formulations of unified theories at different directions in
the moduli space are being used to derive novel features than could not 
be seen, easily, by the use of another unified theory.      
Thus, it appears that problems like that of the correct prediction of 
Newton constant may be solved in four dimensional compactifications, if
M-theory is used \cite{wica}. On the other hand,
F-theory compactifications \cite{mova1,mova2}
are used to examine the appearance of non-perturbative gauge symmetries
for the heterotic string. The common principle in the examination of 
consistency of the compactifications involved is that they all correctly
reproduce, via duality, the weak coupling expansion of the heterotic 
string in four dimensions. 
On the other hand, 4D ${\cal N}=1$ perturbative heterotic string vacua 
and their eleven dimensional 
compactification extensions in four dimensions are considered to be 
phenomenologically promising and furthermore possess 
a lagrangian description contrary to the existing F-theory formulation.

Howerer, the $N=1$ 4D effective supergravity vacua of its low energy 
modes are defined in terms of 
three quantities, namely the K\"{a}hler potential K, the superpotential W, 
and the gauge kinetic function f, which appear as too many. Instead we would 
prefer to have a theory that everything could be defined in terms of only 
one quantity. Is there such a theory? The answer is yes. 
In 4D ${\cal N}=2$ effective low energy supergravity theories, the 
vector 
multiplet sector of the theory, in its Coulomb phase, is defined in terms of 
one quantity, the holomorphic prepotential f. 

However, there is a general weakness in the way we take into account
quantum corrections. 
That is the quantum theory when loop corrections are taken, or not, into
account is not defined so that it is manifestly $Sp(4,Z)$ invariant.
Let us explain this in more detail. 
Take for example the four dimensional compactifications of 
the heterotic string on a $K_3 \times T^2$. The 4D theory has ${\cal N}=2$ 
supersymmetry and originates from further toroidally compactified
the ${\cal N}=1$ $D=6$ heterotic vacua which in turn are dual to F-theory
compactified on a Calabi-Yau 3-fold on an Hirzebrush $F_n$ surface base. 
The vector multiplet 
effective action, all the couplings  in the effective lagrangian, 
is fixed completely, in perturbation
theory, by the knowledge \footnote{The equations which determine directly
the one loop correction to the prepotential $\cal F$ of any ${\cal N}=2$
$D=4$ 
heterotic vacuum for rank three (S-T), and four (S-T-U) models were given 
in \cite{kokos}.} of the holomorphic prepotential ${\cal F}$. In turn,
the K\"ahler potential is defined via the use of the special geometry 
by the holomorphic symplectic vectors $\Omega= (X^I(M^I), F_I(X))$, 
dependent on the moduli fields $M^I$, as
\begin{eqnarray}
K= -\log (-i \Omega^{\dagger} \left( \begin{array}{cc}
0&1\\-1&0
\end{array} \right) \Omega ) =
\log(i {\bar X}^I F_I -i X^I{\bar F}^I),\;\;I=0,\dots,n.
\label{koukos1}
\end{eqnarray}
The target-space duality transformations act on the space of those 
vectors
as $Sp(2n+2, R)$ transformations on the period vector $\Omega$.
Alternatively K can be expressed \cite{mastie} in terms of the T, U 
neutral moduli and the Wilson lines B, C of the $T^2$ torus as
\begin{equation}
K= \log[(T +{\bar T})(U +{\bar U}) - (B +{\bar C})({\bar B}+ C)]=
det(M - M^{\dagger}),  
\label{koukos2}
\end{equation}
where 
\begin{equation}
M = \left( \begin{array}{cc}T&B\\
-C&U \end{array}
\right).
\label{koukos3}
\end{equation} 
In the latter case the effective theory of light modes is invariant,
if we ignore the gravitational sector contribution of the dilaton
and graviphoton, due to the presence of the 
discrete shifts in the theta angle at the quantum level, under the 
target space modular group $Sp(2r;Z)$. Here r is 
the number of moduli. In our case $r=2$ and $Sp(2r;Z)$ acts as
\begin{equation}
M \rightarrow \left( \begin{array}{cc}a&b\\
c&d \end{array}\right)M ,\;\;\;\;\; \left( \begin{array}{cc}a&b\\
c&d \end{array}\right) \in Sp(4; Z).
\label{koukos4}
\end{equation}
However, the calculation of the quantum corrections e.g the one loop
corrections to the gauge or gravitational couplings, the ${\cal N}=2$ 
prepotential, are determined in terms of the basis for modular forms
for the group $SL(2,Z)$ and the $SL(2,Z)$ j-invariant and not those 
associated with $Sp(4,Z)$.

The question that now arises is if the  
theories involved in the compactifications of the five 
perturbative string theories, including M, F-theories, with quantum 
effects may be defined so that they are manifestly $Sp(4,Z)$ invariant? 
Putting the question differently, 
can we express the quantum corrections in terms of $Sp(4,Z)$ entities 
like the basis \footnote{The latter entities are made of  
the usual modular forms for the $SL(2,Z)$ modular group $E_4$, $E_6$ and
the cusp forms ${\cal C}_{10}$, ${\cal C}_{12}$.}
of modular forms for the group $Sp(4,Z)$, making the $Sp(4,Z)$ invariance 
manifest? Note that we wish to define the different unified theories in 
terms of Riemann surfaces having $Sp(4,Z)$ invariance.  
Obviously, the answer to our question is negative. Apart from some 
exceptions unified theories are defined with no manifest $Sp(4,Z)$
 invariance. 
Previous attempts of description of quantum effects where the
ring of modular forms for the $Sp(4,Z)$ is used are evident in some works.
In \cite{mastie} the moduli dependence of the one loop corrections to the
gauge coupling constants in 4D ${\cal N}=1$ orbifold compactifications of 
the heterotic string was translated from the 
language of the basis of modular forms for the $SL(2,Z)$
on the basis of modular forms for $Sp(4,Z)$. In \cite{verlin} 
the calculation of the degeneracy of $N=4$ BPS states was defined 
in terms of genus 2 theta functions.       
However, an apparent question remains on those approaches. Namely, which
is the Riemann surface where the vector moduli of the $SO(3,2)$ T$^2$ torus 
live? 
The latter question
is meaningful both in the context of heterotic string \cite{mastie} and
in the context of F-theory.

Take for example the F-theory/heterotic duality in 
8-dimensions \cite{mova1,vafa1}. Here,
compactifications of F-theory on a $K_3$ which
admits an elliptic fibration with a section is on the same moduli 
space $SO(18,2;{\bf Z}) \backslash SO(18,2)/{{SO(18)} \times SO(2)}$,
as the $E_8 \times E_8$ heterotic string on a $T^2$ torus. 
The elliptic fibration with a section, represented by 
defining a torus with a ${\P}^1$ base, represents the F theory dual 
to the $E_8 \times E_8$ heterotic string are given by a two parameter
$(\alpha, \beta)$ family with base coordinate z, as 
\begin{equation}
y^2=x^3 + \alpha z^4 x + (z^5 + \beta z^6 + z^7).
\label{koukos5} 
\end{equation}
Because the torus is a genus one curve 
the map between the parameters $(\alpha, \beta)$ and the complex
structure parameters of the $T^2$ torus, namely $(\tau, \rho)$,
is expressed in terms of the j-invariant, the modular function for the 
torus \cite{cclm}. As a result at the quantum level the basic 
quantities of the heterotic string are defined in terms of the 
inherited F-theory j-invariant. 
Note that at the large $\rho$ limit the complex structure $\tau$
of the $T^2$ of the heterotic string is identified with the complex 
structure of the elliptic fiber (\ref{koukos5}).
It is the presence of the j-invariant that signals the lacking $Sp(4,Z)$
invariance which is eminent through all the F-theory formulation.

There is another area however, when the question of the Riemann
surface with $Sp(4,Z)$ invariance is meaningful.
Take for example ${\cal N}=2$ supersymmetric Yang-Mills. 
At the Coulomb phase in four dimensions
the moduli space of the r vector multiplets, when charged hypermultiplet 
matter is not present, is invariant under $Sp(4r;Z)$.
The ${\cal N}=2$ $SU(r+1)$theory is associated with genus r Riemann 
surfaces, e.g 
the ${\cal N}=2$ $SU(2)$ theory is described by a genus 1 surface,
the ${\cal N}=2$ $SU(3)$ with
genus 2 and so on. Uniqueness and universality is lost as for $r>1$, the 
theory instead
of being simpler, as the gauge group increases and the number of Wilson 
lines that break the initial "observable" E$_8$ gauge group decreases,
it is defined on even higher genus surfaces. Instead we require in this 
work that we want   
to describe 8D F-theory realizations in terms of double covers of 
$K_3$ fibrations that are branched always along a fixed form plane
curve,
the double sextic. In the latter sence we always work with the universal
form sextic curve.
 
In this work, we will present a Riemann surface that possess $Sp(4,Z)$ 
invariance that we will argue is connected to the 4D heterotic string.
In addition, we will present F-theory solutions in eight dimensions
whose six dimensional versions, that we hope to address in a future work,
may be connected directly to the $Sp(4,Z)$ Riemann surfaces.

In section 2, we describe the connection of the Mordell-Weyl group
to the existence of Weierstrass form of rational elliptic surfaces with
a section.  
We give explicitly the derivation of the Weierstrass models for the 
various forms of torsion subgroups. These Weierstrass models as rational
elliptic surfaces will be used in the next section
to understand the equivalence between $K_3$ surfaces admitting
elliptic fibration and maximizing sextics at the 
degeneration limit of the F-theory/heterotic duality limit.  
In section 3, we describe the 
representation of K$_3$ surfaces as double covers onto $\P^2$  branched 
along
a plane sextic and its connection to 8D F-theory/heterotic duality.
In addition, we explain the correspondence of the plane sextics to the 
extremal elliptic fibrations. 
In section 4, we discuss the Riemann surfaces, appearing in genus two 
and are connected to the calculations of moduli dependence of the one
loop gauge couplings for non-vanishing background fields in $N=1$ four 
dimensional heterotic $(0,2)$ string compactifications, namely the binary 
sextics.   
In section 5, we present our results for the ${\cal N}=2$ 4D 
heterotic theories
which exhibit target space modular group ${\Gamma_o(2)}_T \times 
{\Gamma_o(2)}_U$. These theories may come from further compactification 
on a K$_3$ of 8D F-theory compactifications
with ${\Z}_2$ torsion subgroup or from toroidal compactification of the 6D 
F-theory on a Calabi-Yau 3-fold over a 2D base.

\section{Rational points on Elliptic Curves}

In this section we will explain the appearance of the Mordell-Weyl 
group in 8-dimensional compactifications of F-theory, realised on 
elliptically fibered $K_3$ surfaces.
We will particularly explore, one side of the iceberg, namely
representations of $K_3$ surfaces which are realised 
on genus one curves, those admitting elliptic fibration with a section. 
We are particularly interested in examining the
degeneration limit of the F-theory/heterotic duality\cite{vafa1} where  
the $K_3$ surface breaks into two rational elliptic surfaces.
The other side of the iceberg is the representation of $K_3$ surfaces 
admitting double covers of $\P^2$ along a plane sextic
and their projective realizations and it will be treated in the next 
section .

Let me start  with a few definitions. The Weierstrass form comes from
the most general cubic that can be written in $\P^2$ coordinates as  
\begin{eqnarray}
F(x,y,w) = c_{yyy}y^3 + c_{xyy} xy^2 + c_{xxy} x^2 y+c_{yyw} y^2 w +
c_{xyw} xyw + c_{yww} yw^2 +c_{xxx} x^3 + &\nonumber\\
c_{xxw} x^2 w + c_{xww} xw^2
+ c_{www} w^3 .&
\label{eksi2}
\end{eqnarray}

When a number of conditions is applied to (\ref{eksi2}) it can always 
be reduced in the general projective\footnote{
The subscripts of the coefficients denote the homogeneity of the 
corresponding term under a change of variables.}
Weierstrass form in $\P^2$ which reads
\begin{equation}
y^2 w + a_1 xyw + a_3 y w^2=x^3 + a_2 x^2 w + a_4 xw^2 +a_6 w^3.
\label{eksi1}
\end{equation}
These conditions can be summarized as follows :

$\bullet$ Demanding the curve (\ref{eksi2}) to pass 
through $(x,y,w)=(0,1,0)$ we get $c_{yyy}=0$.

$\bullet$ Having (0,1,0) as a non-singular point gives us $c_{yyw} \neq 0$
or 
$c_{xyy}  \neq 0$.

$\bullet$ Having the curve  $f(x,y,w) =w$ as a tangent line at infinity
point (0,1,0) we get one more condition $c_{xyy}=0$ and from the previous 
condition
$c_{yyw} \neq 0$.
  
$\bullet$Lastly because a non-singular point P of F is a flex or inflection
point
if the intersection 
multiplicity of the tangent line of F at P is greater or equal to 3, the curve 
 (\ref{eksi2}) can have an inflection point at (0,1,0) if 
$c_{xxy}=0$  and $c_{xxx} \neq 0$. 

Under these conditions the cubic takes the form
\begin{equation}
F=c_{xxx} y^2w + c_{xyw} xyw + c_{yww} yw^2 + c_{xxx} x^3 + c_{xxw}x^2 w
+ c_{xww} xw^2 + c_{www}w^3.
\label{akle}
\end{equation}
This form is known as Weierstrass form. By setting $w=1$ we can put the 
cubic in the affine Weierstrass form as 
\begin{equation}
y^2 + a_1 xy + a_3 y = x^3 + a_2 x^2 + a_4 x + a_6.
\label{akle1}
\end{equation}
The subscripts in the coefficients of the non-singular cubic indicate the
degree of homogeneity under a certain change of variables.
Note that (\ref{akle1}) is singular only if $a_3=a_4=0$.

As usual, compactifications of heterotic string on $T^2$, in the context of 
its duality with F-theory 
on a $K_3$ surface in
8 dimensions, are realized in terms of an elliptic curve 
over\footnote{Note that the affine plane
$k^2={(x,y )}$ has a standard one to one embedding into $\P_2(k)$. 
In turn , 
$\P_2(k)$ is defined as the quotient of $(x,y,w) \in (k^3 - {(0,0,0))}$. }
a field  k, namely as a non-singular cubic that is in a Weierstrass 
form. 

An elliptic curve over a the field of rationals Q is defined as a 
non-singular cubic in 
Weierstrass form with rational coefficients. The elliptic curve over Q is 
realized
by completing the square in the Weierstrass form followed by a change of
 variables in y to get $y^2=R(x)$, where 
R(x) a cubic in degree 3 with distinct roots.
In the latter form, namely (\ref{akle1}) the inflection point is mapped to 
the point at infinity ${\cal O}=(0,1,0)$ such that it becomes the line at 
infinity.
For the non-singular cubic curve, the existence of a specified point 
$\cal O$, defines a group operation on the curve associating the points on 
the curve
with an abelian group, the Mordell-Weyl group (MW), with $\cal O$ its
identity element.   
 What the MW group theorem says is that the group of rational
points of an elliptic curve over Q, E(Q),  is finitely generated. 
The MW group can be written as $E(Q) \approx Z^r \oplus \Phi$ , where $\Phi$
is
a finite abelian group known as the torsion subroup. The integer r is the 
rank of E(Q).
Given now the affine cubic in (\ref{akle1}) and the definition of the 
MW 
group we need to know how it is possible to construct for a given MW group
its Weierstrass form. 
The Weierstrass form for the diffferent choises of the MW group may be 
associated with the appearance of the non-simply connected gauge groups 
\cite{asmo} in 8-dimensional compactifications of F-theory on a K$_3$ 
surface
at its degeneration limit in the next section.
We note that the torsion subgroup
F of an elliptic curve when the field of integers k is equal to the field 
of rational functions becomes the group of sections.

Before addresing the Weierstrass construction,  
let me give first few definitions concerning the group formation on 
elliptic curves that will help us to understand the procedure.

$\bullet$ Given an initial point $\cal O$, we define the group
law on non-singular cubics in terms of the identity element fixed
at ${\cal O}=(x,y,w)=(0,1,0)$. In addition, we 
define $R=PQ=P \cdot Q$ as the line element between two different 
points P, Q on the 
curve. The addition law for a point in the curve is defined as the 
multiplication
$P+Q= {\cal O} \cdot PQ$. This operation makes the points on the curve to
form an abelian group with $\cal O$ as the identity element, 
$P + {\cal O}= {\cal O} \cdot P{\cal O} = P$.
Negatives 
are defined by first setting ${\cal O}{\cal O}$ as 
the third point on the line tangent at ${\cal O}$.
By definition $-P \stackrel{def}{=}{\cal O}{\cal O} P$. 
This means
$-P \stackrel{def}{=} (OO) \cdot P $. So  
$P + (-P) = {\cal O} \cdot P(-P) = {\cal O} \cdot {\cal O}{\cal O}= 
{\cal O}$.
Note for the elliptic curve over the field of rationals we define
${\cal O}{\cal O} = {\cal O}$ as the inflection point. In addition, 
for a general point $P=(x_o, y_o)$ on the 
elliptic curve (\ref{akle1})
the inverse $-P=(x_o, -y_o -a_1 x_o - a_3)$. We can easily see that for
example when
\begin{equation}
a_1=a_3=0,\;\; -P=(x_o, -y_o).
\label{akleo1}
\end{equation}

Next let us look at the determination of the Weierstrass form
of the elliptic fibrations admitting torsion points using the affine 
form (\ref{akle1}).
For the study of the Mordell-Weyl group it is enough to  examine the
 behaviour of the elliptic curve (\ref{akle1}) at 
at the point P=(0,0). The latter means $a_6=0$ and eqn. (\ref{akle1})
becomes
\begin{equation}
y^2 + a_1 xy + a_3 y =x^3 + a_2 x^2 + a_4 x.
\label{aret}
\end{equation}
Now what happens is that E(k) becomes an abelian group with 
${\cal O}=(0,1,0)$
the point at infinity as identity element of the group law on the elliptic 
curve.
The condition that the torsion subgroup $\Phi \cong {\Z}_2$ is equivalent to 
the 
statement that $P + P ={\cal O}$ (or $P = -P$ ). 
In other words P is of order two if the tangent is $\infty$ at P, that 
means vertical tangency.
Since the point at infinity is non-singular, to study singular points,
as we said already we translate the points on the curve at the point $(0,0)$.
That had set $a_6=0$. Note that eqn. (\ref{akle1}) is singular only when 
$a_3=a_4=0$.
By taking now differentials in eqn. (\ref{akle1}) we deduce that the
coefficient
of dy equals zero at P. This means that $a_3=0$. 
Now the Weierstrass form of the elliptic curve becomes
\begin{eqnarray}
\bullet\;\;\;\;\;\;\Phi \cong {\Z}_2,\;\;\;\;\;\;\;\; 
y^2 +a_1 xy = x^3 + a_2 x^2 +a_4 x .\;\;\;\;\;\;\;\;\;\;\;\;
\;\;\;\;\;\;\;\;\;\;\;\;\;\;\;\;\;\;\;\;\;.
\label{akleo2}
\end{eqnarray}
In the special case that $a_1=0$ (\ref{akleo2}) gives the same 
Weierstrass equation as that appearing in \cite{kubert,asmo} . 
In general for an elliptic curve over the field $k=Q$, $\Phi$ must be one 
of the following fifteen groups\cite{kubert} 
\begin{eqnarray}
{\Z}, {\Z}_2,\;{\Z}_3,\; \dots {\Z}_9,\; {\Z}_{10},\; {\Z}_{12},\; 
{\Z}_2 \oplus {\Z}_2,\;
{\Z}_4 \oplus {\Z}_2,\;
{\Z}_6 \oplus {\Z}_2,\; {\Z}_8 \oplus {\Z}_2
\label{akleo3}
\end{eqnarray}

For an elliptic curve over the complex numbers $k=C$ there are in addition 
four more possibilities.
They are listed by Cox and Parry\cite{cox} as
\begin{eqnarray}
\begin{array}{cccc}
 {\Z}_3 \oplus {\Z}_3,\; {\Z}_6 \oplus {\Z}_3,\; {\Z}_4 \oplus {\Z}_4,\;
{\Z}_5 \oplus {\Z}_5.
\end{array}
\label{akleo4}
\end{eqnarray}   
In the case where the 
points of the elliptic surface
over $C(t)$, are rationals, the rational 
elliptic curve can be written as an elliptic surface with a section S, a 
Jacobian\cite{mipe}. 
Note that the Weierstrass forms that we examine in this paper 
are associated with standard compactifications of F-theory 
\cite{vafa1,mova1,mova2} in 8 
dimensions where the antisymmetric B-field is zero\footnote{Non-standard 
compactifications of F-theory where the latter field is turned on
where considered in \cite{bers,berg}.}.
As the methods discussing other possibilities for the torsion 
subgroup $\Phi$\footnote{More details can
be found at \cite{knapp}.} are not present in string theory literature
we can briefly discuss them here.   
If we examine higher orders in the torsion subgroup $F$ we need the 
coordinates of the point 2P. At the tangent line at a general point
P=(x,y) on the cubic (\ref{akle1}) the coordinates of the point 2P are 
given by 
\begin{equation}
x(2P) = \frac{x^4 -b_4 x^2 -2b_6 x -b_8}{4x^3 +b_2 x^2 + 2b_4 x+b_6},
\label{aklie1}
\end{equation}
where 
\begin{eqnarray}
b_2 = a_1^2+ 4a_2,\;\;\;\;\;\;\;\;\;\;\;\; b_4 = 2a_4 + a_1 a_3,&
b_6=a_3^2 + 4a_6 ,&\nonumber\\
b_8=a_1^2 a_6+ 4a_2 a_6 -a_1 a_2 a_4 + a_2 a_3^2 -a_4^2.&&
\label{aklier1}
\end{eqnarray}
Let us examine its consequences. For the singular 
point $P= (0,0)$ eqn. (\ref{akleo1}) gives 
\begin{equation}
-P = (0,a_3) \;\;\;\;\;\;\;2P=(-a_2, a_1 a_2 -a_3)
\label{riri1} 
\end{equation}
Lets us see now how torsion points of order 
three, $\Phi \cong Z_3$  and higher are embedded in the general 
equation (\ref{akle1}) of the elliptic curve.
Let us now make the following isomorphic\footnote{
Changes of variables in a Weierstrass equation in the form
$x \rightarrow u^2 x^{\prime} + r$, $y \rightarrow u^3 y^{\prime}
+ s u^2 x^{\prime} + t$ fix the point $(0,1,0)$ and carry the line 
w=0 at the same line. The latter is important
since the line $w=0$ is the line at infinity. }
change of variables in eqn. (\ref{aret}).
\begin{equation}
(x,y) \rightarrow (x^{\prime}, y^{\prime} + a_3^{-1} a_4 x^{\prime})
\label{aret1}
\end{equation}
gives us that 
\begin{equation}
y^2 + a_1 xy + a_3 y = x^3 + a_2 x^2.
\label{opili}
\end{equation}
Torsion of order three means $2P + P = {\cal O}$. In other words
$2P =-P$, $-P=(0, a_3)$ and $a_2 =0$. 
Thus for
\begin{eqnarray}
\bullet\;\;\;\;\;\;\;\;\;\;\;\;\;\;&\Phi\cong {\Z}_3,&\;\;\;\;\;\;\;\;
\;\;\;\;\;\;\;\;\;\;\;\;\;\;\;\;\;\;\;\;\;\;
\;\;\;\;\;y^2 + a_1 xy + a_3 xy 
= x^3 .\;\;\;\;\;\;\;\;\;\;\;\;\;\;\;\;\;\;\;\;\;\;\;\;\;\;\;\;
\label{opili1}
\end{eqnarray}

Lets us now return back into eqn. (\ref{opili}). Making another change 
of variables, to an isomorphic curve, in the form
\begin{equation}
(x,y) \rightarrow (\frac{x^{\prime}}{u^2}, \frac{y^{\prime}}{u^3}) , 
\label{were1}
\end{equation}
we get that the point P=(0,0) remains fixed and (\ref{opili})  
is transformed into 
\begin{equation}
y^2 + a_3^{-1} a_1 a_2 xy + a_3^{-2}a_2^3 y =x^3 + a_3^{-2}a_2^3 x^2 .
\label{iopu12}
\end{equation}
or in more elegant form into
\begin{eqnarray}
\;\;\;\;\;\;\;\;\;\;\;\;\;\;\;\;\;\;\;\;\;\;\;\;\;\;\;\;\;\;
\;\;\;\;\;\;\;\;\;\;\;\;\; y^2 +(1-c) xy-by = x^3 -b x^2, \;\;\;
\;\;\;\;\;\;\;\;\;\;\;\;\;\;\;\;\;\;\;\;\;
\label{akleo22}
\end{eqnarray}
where $b= - a_3^{-2} a_2^3$ and $c=1-a_3^{-1} a_1 a_2$.
The Weierstrass form of eqn.(\ref{akleo22}) is well known 
in mathematics literature as Tate normal form. 
Its discriminant is given by 
\begin{equation}
\bigtriangleup(b,c) = (1-c)^4 b^3 -(1-c)^3 b^3 -8(1-c)^2 b^4 + 36(1-c) b^4 
+ 16 b^5 -27 b^4
\label{iopu14}
\end{equation}
and its connection to the usual from F-theory considerations discriminant 
$\bigtriangleup= 4 \alpha^4 + 27 \beta^2$ may be obvious.

We can now look at the derivation of the Weierstrass form for higher than
${\Z}_3$ torsion subgroups. 
Looking at (\ref{iopu12}) and repeating the previous procedure
that was applied at the ${\Z}_2$ case, namely requiring the tangent at 
the point P to be tangent, we get
\begin{eqnarray}
\begin{array}{ccc}
P=(0,0),&2P =(b,bc),&3P=(c,b-c), \\
-P=(0,b),&-2P=(b,0),&-3P=(c,c^2).\\
\end{array}
\label{relat}
\end{eqnarray}
By taking subsequently
\newline
$3P =-P$ with $c=0$,
\begin{eqnarray}
\bullet\;&\Phi\cong {\Z}_4 ,&\;\;\;\;\;\;\;\;\;\;\;\;\;\;\;\;\;\;\;\;\;\;\;
y^2 + xy -b y = x^3 -bx^2,\;\;\;\;\;\;\;\;\;\;\;\;\;\;\;
\;\;\;\;\;\;\;\;\;\;\;\;\;\nonumber\\
&\bigtriangleup =\bigtriangleup(b,o) =b^4 + 16b^5.&
\label{opili24}
\end{eqnarray}
$3P=-2P$ with $b=c$,
\begin{eqnarray}
\bullet\;\;\;\;\;\;\;\;\;&\Phi\cong {\Z}_5,&\;\;\;\;\;\;\;\;\;\;\;\;\;\;\;\;\;\;\;
y^2 + (1-c) xy -b y = x^3 -bx^2,\;\;\;\;\;\;\;\;\;\;\;
\;\;\;\;\;\;\;\;\;\;\;\;\;\;\;\;\;\;\;\;\;\;\;\nonumber\\
\;\;\;\;\;\;\;\;\;\;\;&\bigtriangleup =\bigtriangleup (c,c),&
\label{opili25}
\end{eqnarray}
$3P=-3P$ with $b=c^2 +c$.
\begin{eqnarray}
\bullet\;\;\;\;\;\;\;\;\;\;\;\;&\Phi\cong {\Z}_6,&\;\;\;\;\;\;
\;\;\;\;\;\;\;\;\;\;\;\;\;\;\;\;\;\;\;\;\;
y^2 + (1-c) xy -(c^2 +c) y = x^3 -(c^2 +c)x^2,\;\;\;\;\;\;\;\;\;
\nonumber\\
\;\;\;\;&&\bigtriangleup =\bigtriangleup (c^2+c, c).
\label{opili26}
\end{eqnarray}
There are two remaining cases of Weierstrass forms representing 
rational elliptic surfaces where the torsion is a subgroup
of $Z_{2n} \times {\Z}_2$. In those cases the rational elliptic 
surface is associated with
\begin{eqnarray}
\bullet\;\;\;\;\;\;\;\;\;\;\;&\Phi \cong {\Z}_2 \oplus {\Z}_2,&\;\;\;\;\;\;\;\;\;
\;\;\;\;\;\;\;\;\;\;\;\;\;\;\;\;\;\;\;\;\;\;\;\;\;\;\;\;\;\;\;\;\;\;\;\;\;\;
y^2 =x(x-\beta)(x-\gamma),\;\;\;\;\;\;\;\;\;\;\;\;\;\;\;\;\;\;\;
\label{ert1}
\end{eqnarray} 
with $\beta,\;\gamma\;\in\;\Z$.
\begin{eqnarray}
\bullet\;\;\;\;\;\;\;\;\;\;\;&\Phi\cong  {\Z}_4 \oplus {\Z}_2,&\;\;\;
\;\;\;\;\;\;\;\;\;\;\;\;\;\;\;
\;\;\;\;\;\;\;\;\;\;\;\;\;\;\;\;\;\;\;\;\;\;\;\;\;\;\; y^2=x(x+
\zeta^2)(x+ \lambda^2).\;\;\;\;\;\;\;\;\;\;\;\;\;\;\;\;\;\;
\label{ert2}
\end{eqnarray}
Equation (\ref{ert1}) appears in this form after the use of
specific theorems while (\ref{ert2}) comes by demanding
that one of the points $(x_o,0)$, with $x_o=0, \beta, \gamma$
 in (\ref{ert1}) must be the double of another number in order for 
the ${\Z}_2 \oplus {\Z}_2$ to flow at ${\Z}_4 \oplus {\Z}_2$.
That means that $\beta=\zeta^2$, $\gamma=\lambda^2$ in (\ref{ert2}).
The Weierstrass representations of the elliptic fibrations with a
section in eqn.'s
(\ref{akleo2}), (\ref{opili1}), (\ref{opili24}), (\ref{opili25}),
(\ref{opili26}), (\ref{ert1}), (\ref{ert2}) 
are all relevant at the degeneration limit of F-theory on the elliptic
$K_3$ studied in the next section.  

\section{Eight-dimensional compactifications of F-theory as 
$K_3$ fibrations}

\subsection{\em Why double covers?}

In this section we examine 8 dimensional compactifications of F-theory 
and examine the heterotic/duality map\cite{vafa1,mova1}
at the stable degeneration limit\cite{fmowi,pamo}.
According to this duality map, F-theory on a $K_3$ surface
admitting an elliptic fibration with a section is on the same moduli
space as the heterotic string on a $T^2$ torus.  
Take for example the $E_8 \times E_8$ heterotic string. 
Then at the limit that the $T^2$ torus is large, the $K_3$ surface
degenerates into a variety that is made from the union of two intersecting
rational elliptic surfaces $S_1$, $S_2$ intersecting along an elliptic
curve $E^{**}$. At this limit the J-invariant of the F-theory elliptic
curve $E^{**}$ coincides with the J-invariant of the heterotic elliptic 
curve \cite{pamo} representing the $T^2$ torus as\footnote{the double
cover of the complex plane x} 
\begin{equation}
y^2=x^3 + a x + b.
\label{koriko1}
\end{equation}

On the contrary the $K_3$ $S_F$ compactification of F-theory
is represented as an elliptic fibration over a $\P^1$ base, namely
as a map $\pi :S_F \rightarrow \P^1$. 
This constitutes a form of violation of "double cover parity"
signalling different treatment of the bases of the duality pairs.
We demand that  
the violation of double cover parity,
of the F-theory/heterotic duality map can be corrected by considering
generating the $K_3$ surface from double covers. The latter {\em
happens}, in our case, by considering
double covers of $K_3$ onto $\P^2$, the so called double sextics. 
It is known \cite{griha} that double covers of $K_3$ onto $\P^2$ are branched 
along a plane sextic curve. That idea will constitute our main tool in this 
section.
The difference between double covers onto $\P^2$ and 
$\P^1$ fibrations,
between the two dual sides, can intuitively be explained 
by the following correspondence.

Consider the cubic defined by
\begin{equation}
xy^2 =z(x-z)(x-\lambda z).
\label{asdf1}
\end{equation}
Then two generic members of the cubic pencil\footnote{Note that
the term pencil is defined to be the dimension of the projective space
that parametrizes eqn.(\ref{asdf1}). Namely one in our case.}
may be described as cubic curves with common flexes at $(0,1,0)$
and common tangents at $(0,1,0)$, $(1,0,1)$ and $(\lambda, 0, 1)$.    
By blowing up the base points one can obtains 
the quotient of $E \times \P^1$ under the involution 
\begin{equation}
(y, z) \stackrel{\sigma}{\rightarrow} (-y,-z), 
\label{akll1}
\end{equation}
where E is the elliptic curve connected to $(0,1,\infty, \lambda)$. 
The notation $(y, z)$ refers to the elements of $E \times \P^1$.
In turn, two members of the cubic pencil together with its two
degenerate fibers calculate four points on $\P^1$ and thus an elliptic 
curve $E^{\prime}$. As a result the double cover of the rational elliptic 
fibration along two members has apriori its double cover given
by the abelian surface $E \times E^{\prime}$. 
If we can connect in some way the double cover of some form of
abelian surface
in some form of F-theory compactifications on K$_3$ as a double cover we
are done. 
\newline
{\em Maximizing sextics and singular $K_3$ surfaces}

The way to proceed is to connect the double sextics to
singular $K_3$ surfaces\cite{shioino}, e.g $K_3$ surfaces for which
the Picard number $\rho = h^{1,1}$. Since every singular $K_3$ surface is
the double cover of a Kummer surface thay can be connected to 
compactifications involving orbifold limits of $K_3$ and orientifolds.     
Let us first give some definitions.
 The lattice associated to the
latter has rank 
$\rho_{NS} =20$ and signature $(+ 1, (-1)^{19})$.
In general to singular $K_3$ surfaces one associates\cite{shioino}
the Picard number 
\begin{equation}
\rho^{sin}_{K_3}= 2 + \sum_{\nu =1}^{k} \mu(E_{\nu}) + rk|\Phi|,
\label{akler1}
\end{equation}
where $|\Phi|$ is the order of the group of sections $\Phi$ that is the
Mordell-Weyl
group and r its rank. The quantity $\mu(E_{\nu})$ is associated to the 
number of components of the set $E_{\nu}$ of the singular fibers.
Similarly, when a curve X is the double cover of Y branched 
along a curve C, then 
\begin{equation}
\rho(X) \geq \rho(Y) + \sigma(C).
\label{asdf11}
\end{equation}
For a plane sextic curve C coming as a double cover onto $\P^2$, 
$0 \leq \sigma \leq 19$. So by comparison with (\ref{akler1})
we can notice  that
the index\footnote{When we minimally resolve a singularity $x_n$, n linearly
independent rational
curves in the Neron-Severi group appear.  
Then for a curve C the index is defined as the sum of the
subindices n of all its the simple singularities.}
 $\sigma(C)$ associated to the sextic has a similar role
as those of the singular fibers of the singular $K_3$ surfaces. 
In fact we will see, in the rest of the section, in the cases of interest
in 8 dimensional
compactifications of F-theory/heterotic duality map, $K_3$ surfaces
admitting
elliptic fibrations with a section and $K_3$ surfaces admitting double
covers on to $\P^2$ coincide.
The latter list of $K_3$ fibrations has been worked out in \cite{ena}. 
So by considering double covers onto the base manifold on both sides 
of the 8-dimensional F-theory/heterotic duality map we recover naturality 
in the treatment of the base. 

A sextic is called maximizing if its index is maximal, $\sigma=19$. 
A a result from (\ref{asdf1}) it follows that the maximizing sextic C is
a singular $K_3$ surface.
If the index    
$\sigma(S) =19$ the singularities of the sextic
are in the form
$a_1$, $d_{2n}$, $e_7$ and $e_8$. In this case the sextic is called 
supermaximizing and the number of the rational components in this case
is less than ten. That can be proved by a simple calculation of the 
Euler number.  
We remind that these are singularities of the branch curve and not its
double cover. The double cover has as usual A, D, E singularities.
This suggests that in the general case $K_3$ surfaces that are acted by 
involutions not having fixed points have singular fibers with at least
ten components. When we have ten components those components are rational. 
In fact the following table will assist us to our work later.
\begin{center}
\begin{tabular}{|c|c|c|c|c|c|c|} \hline
fiber-type& singularity-type & $\epsilon$ &  $\delta$ &
d&order of torsion group& double cover \\ \hline\hline
$I_n$ & $A_{n-1}$ &n & n-1 &n & $\infty$ & $I_{2n}$ \\\hline
$I^{*}_{n}$ & $D_{4+n}$ & n+6 &n+4&4 &4 &$I_{2n}$ \\ \hline
II&none&2&0&1&1&IV \\ \hline
III&$A_1$&3&1&2&2&$I_o^{*}$ \\ \hline
IV&$A_2$&4&2&3&3&$IV^{*}$\\ \hline
$II^{*}$&$E_8$&10&8&1&1&$IV^{*}$ \\ \hline
$III^{*}$&$E_7$&9&7&2&2&$I_o^{*}$ \\ \hline
$IV^{*}$&$E_6$&8&6&3&3&IV \\ \hline
\end{tabular}
\end{center}
\begin{center}
Here, the Kodaira fiber are placed against their components $\delta$,
Euler number $\epsilon$, discriminant and fiber of the double cover.
\end{center}

\subsection{\em Elliptic fibrations with involutions}

The double cover consists of an elliptic fibration $\pi : Y ->{\P}^1$
and an involution $\sigma : z -> -z$ that respects the fibration.
In the general case that Y is an elliptic $K_3$ surface
we can distinguish two general cases.
The case of isolated fixed points and the case of no isolated fixed
points. In the first case the are eight
 fixed points and the quotient by the involution X is a $K_3$ surface. 
That is the case that is considered in
 studying the CHL vacuum in \cite{bers}.
In the second case there are three possibilities:
a) X is a rational surface with the fixed locus appears in two components.
The surface Y may be coming from a double sextic, as a union of two cubics.
b) X is a rational surface but the fixed locus appears in one component,
and c) X is an Enriques surface with no fixed locus and the
elliptic fibration has two double fibers.
The point that we follow is to consider elliptic $K_3$ fibrations
$\pi : Y \rightarrow \P^1$, where Y is an elliptic $K_3$ surface, 
and examine the effect of the involutions $\sigma$ that respect the 
fibration. We represent the involution after resolution by
$ X=\tilde{Y/{\sigma}}$ where tilde denotes desingularization.
We will be interested in singular sextics that can give $K_3$ surfaces
after resolution. 
In fact, there are two possibilities that can be realized for the surface X.
The surface X can be either a ruled surface or an elliptic surface.
Those two distinctions are related to the way that the involution $\sigma$
acts on the Y fibration.
We can have X as a rational ruled surface 
only if each fiber is invariant under the involution 
and contains fixed points.
In fact, by blowing down X to a minimal model $X_o$, containing no
exceptional curves of the first kind, that can be the Hirzebrush surface
${\bf F}_4$, we obtain Y, the $K_3$ surface, as a double cover of $X_o$
branched along a
4-section $\Gamma_o =S_{\infty} + T$.  We denoted by $S_{\infty}$ the 
minimal section, $S_{\infty}^2=-4$ and T is a trisection disjoint 
from $S_{\infty}$. 
One can now observe that the index of the elliptic fibration is 
the index of $\Gamma_o$ or T. 
In this case the branch locus of the covering will be a transversal
divisor hitting each rational fiber exactly four times and it will appear
as a sextic.
In this case the rational surface X can be represented as a $\P^2$ 
after afew birational modifications.  
In particular the $K_3$ surface Y with an
extremal fibration is realized as the double covering of a maximizing 
sextic, only if the fibration has three singular fibers of the type
$I_2$, $I_{n}^{*}$, $II^{*}$, $III^{*}$, $III$.
Something that was practically suggested in the previous subsection.
The second case  that we discuss is when X is an elliptic surface.
Now the quotient transposes the fibers 
i.e permutes them and the branch locus is made of two fibers.
In this case the elliptic $K_3$ surface Y may be obtained as a double
sextic from a union of two cubics.

An easy way to get an elliptic fibration is to have a singular point P
on the sextic and consider the elliptic fibration induced by the lines
through it. When done in this way, the elliptic fibration centered at the
 triple
point of a maximizing sextic is extremal. 
In this case there are more possibilities realized for the rational 
surface X as the torsion group can be different thann unity as it was the case
in the standard heterotic/F-theory duality \cite{vafa1,mova1,mova2}. 
The computation of the torsion 
group
$\Phi$ for elliptic fibrations of a maximizing sextic can be determined
 from the relation
\begin{equation}
d(X)= {\Pi}_{\nu} d(F_{\nu})/|\Phi|^2.
\label{were1}
\end{equation}
Here, d denotes the discriminant while 
includes the blown up Kodaira singularities and the product runs over
the configurations of the singular fibers.
Take for example the configuration with $I_9, 3I_1$ fibers.
Applying the formula (\ref{were1}) we get that $d(X)=1$ which must be the 
case
as the
Picard lattice for a rational elliptic surface is the whole of $H^2(X, Z)$
and the lattice is unimodular. 

The results in the case of double covers 
are summarized as follows:
\begin{center}
\begin{tabular}{|c|c|r|} \hline
Cubic fibration & order F &
{$K_3$ fibration}\\ \hline\hline
$II^{*}$ $2I_1$&1& $2II^{*}$ $2I_2$\\ \hline
$II^{*}$ II&1& $2II^{*}$ IV\\ \hline
$III^{*}$ $I_2$ $I_1$&2& $2III^{*}$ $I_4$ $I_2$ \\ \hline
$III^{*}$ III&2&$2III^{*}$ $I_0^{*}$\\ \hline
$I_4^{*}$ $2I_1$&2&$2I_4^{*}$ $2I_2$\\ \hline
$I_9$ $3I_1$ &3&$I_{18}$ $I_2$ 4$I_1$, 2$I_9$ 2$I_2$ 2$I_1$ \\ \hline
$IV^{*}$ $I_3$ $I_1$&3&$2IV^{*}$ $I_6$ $I_2$\\ \hline
$IV^{*}$ IV&3&3$IV^{*}$\\ \hline
$I_8$ $I_2$ 2$I_1$ &4&$I_{16}$ $I_4$ 4$I_1$, $I_{16}$ 3$I_2$ 2$I_1$,
$I_{16}$ 3$I_2$ 2$I_1$, 2$I_8$ $I_4$ $I_2$ 2$I_1$,
2$I_8$ 4$I_2$\\ \hline
$I_2^{*}$ $2I_2$&4&$2I_2^{*}$ $2I_4$\\ \hline
$I_1^{*}$ $I_4$ $I_1$&4& $2I_1$ $I_8$ $I_2$\\ \hline
2$I_5$ $2I_1$&5&2$I_{10}$ $4 I_1$, $I_{10}$ $2I_5$ $I_2$ $2I_1$,
$4I_5$ $2I_2$\\ \hline
$I_6$ $I_3$ $I_2$ $I_1$&6&$I_{12}$ $I_6$ $2I_2$ $2I_1$, $I_{12}$ $I_4$
$2I_3$ $2I_1$, $I_{12}$ $2I_3$ $3I_2$, $3I_6$ $I_4$  $2I_1$,\\ 
&&$3I_6$ $3I_2$, $2I_6$ $I_4$ $2I_3$ $I_2$\\ \hline
$2I_4$ $2I_2$&8& $2I_8$ $4I_2$, $I_8$ $3I_4$ $2I_2$, $6I_4$ \\ \hline
$4I_3$&9&$2I_6$ $I_3$\\ \hline 
\end{tabular}
\end{center}
\begin{center}
Table 2
\end{center}
The first column 
gives us the the Kodaira fibers appearing in the cubic associated with
 the corresponding Mordell-Weyl group order of the group of sections
given in the second column. The third column is the configuration of
allowed Kodaira fibers for the associated $K_3$ fibrations.

\subsection{\em Examples with Point Like Instantons}

We will discuss the case of the creation of an $E_8$ gauge symmetry
in the F-theory/heterotic duality realization in eight 
dimensions\cite{fmowi,pamo} when the $K_3$ surface $S_F$ of F-theory
compactifications is acted by the involution $\sigma: z-> -z$.
The resulting surface X is an elliptic surface. Now the sextic
of the branch locus  is made
from the union of the two cubics. At the limit that the heterotic
torus is large, 
the $K_3$ surface breaks up into two rational surfaces and 
and each rational surface is coming by
taking the double covers branched at two fibers. 
Elliptic fibrations generated in this way always have sections. 
In the case of $E_8 \times E_8$ symmetry on the F-theory side, 
the observed symmetry group $E_8$
is reproduced by the elements of the vanishing cohomology $H_2(S_i,\Z)$,
with i either 1 or 2, of the rational elliptic surface $S_i$.
In geometrical terms the rational elliptic surface represented as a
cubic $C_0$ and a flexed line $L_{\infty}$ forms a cubic pencil
spanned by $C_0$ and $3L_{\infty}$. This has a $II^{*}$ and 
$2I_2$ fibers, an thus generates an unbroken $E_8$ gauge symmetry 
from point like instantons\cite{point,mova1} with no local holonomy.
Taking the double cover of the sextic formed by the two nodal cubics in
the pencil will produce an elliptic fibration of type $2II^{*} 2I_2$.
Amazingly enough it appears that the double cover of the $K_3$
fibration
can create the whole of $E_8 \times E_8$ gauge symmetry. However,
$2II^{*} 2I_2$ is the symmetry of the double cover and does not
correspond to the observed heterotic gauge symmetry.

Another elliptic fibration on the same $K_3$ may be build by 
considering the lines through one of the nodes. Now the intersection of 
the
two cubics\footnote{Remember that a rational elliptic surface 
is the blow up of $\P^2$ at the nine intersection points of two cubic
curves, or the nine basepoints of the cubic pencil.} of order nine will
give rise to an $I_{18}$, the other node will give us a $I_2$ and there
are four tangent lines from the point to the nodal cubic.
In total we get the configuration $I_{18}I_2$$ 4I_1$ with the order of the 
Mordell-Weyl group to be three.
The $Spin(16)/Z_2$ heterotic string comes from the configuration
$I_4^{*} 2I_1$ of the rational elliptic surface of table 2. The
corresponding $K_3$ fibration upon which the involution $\sigma$ is
acting is $2II^{*} IV$. In fact more possibilities are possible
\cite{nanou}.

\section{ Effective string theories from double covers} 

The question that remains from the previous considerations 
is if we can find an effective theory that can be formulated in terms
of double covers and is defined in lower dimensions, e.g four, that may
be defined in the context of F-theory/heterotic duality. The 
heterotic string theory defined in terms of double covers is considered in 
this section. We hope to address the $N=2$ four dimensional
compactification of F-theory in terms of double covers in a future work.
Let us now consider the usual  
F-theory compactification, on a 
Calabi-Yau 3-fold with an $F_n$ base \cite{mova1,mova2} which is on the 
same moduli space as the
heterotic string on a $K_3$ surface. Further compactification on four 
dimensions on both sides on a $T^2$ torus may produce a $N=2$ vector 
multiplet 
effective theory with e.g an $SL(2,Z)_T \times SL(2,Z)_U$ duality 
invariance. 
If there are Wilson
lines on the $T^2$ torus the classical perturbative duality invariance of the 
vector multiplet effective theory becomes $Sp(4,Z)$.  
In this case the formulation of the effective theory of light modes is
consistently described in terms on the genus two Siegel modular forms
which can reproduce consistently the one loop corrections to the 
gauge coupling constants of $N=1$ $(0,2)$ orbifold compactifications  
of the heterotic string\cite{mastie}.
As we said in the introduction it is desirable to have an 
interpretation of the formulation of the heterotic string
compactifications
in terms of particular Riemann surfaces with manifest $Sp(4,Z)$
invariance. Here we will do just that.
The way that we will proceed in our investigation is to first
define the basic elements that define the $N=1$ $(0,2)$ orbifold
compactifications of the heterotic string and then    
describe the derivation of the Riemann surface possessing the
$Sp(4,Z)$ invariance.

For genus $g=1$ the graded ring of modular forms
is generated by the $E_4$, $E_6$ modular forms\footnote{the subscript
denotes the modular weight of the respective modular form.}.
In the genus two case the graded ring is generated by the Siegel modular
forms $E_4$, $E_6$, and the cusp forms $\chi_{10}$, $\chi_{12}$ and
$\chi_{35}$ where the latter are given by the following expressions 
\begin{center}
\begin{eqnarray}
I_4=\sum (\theta_m)^8 ,\nonumber\\
I_6= \sum \sum_{syzygous}\pm (\theta_{m_1} \theta_{m_2} \theta_{m_3})^4
, \nonumber\\
I_{10} = -2^{14} \cdot \chi_{10} = \Pi (\theta_m)^2 ,\nonumber\\
I_{12}=2^{17}3 \cdot \chi_{12}=\sum
(\theta_{m_1} \theta_{m_2} \dots \theta_{m_6})^4 ,\nonumber\\
I_{35}=2^{39}5^3 i \cdot \chi_{35} = (\Pi \theta_m)(\sum_{azygous} \pm
(\theta_{m_1} \theta_{m_2} \theta_{m_3})^{20}),
\label{igu1}
\end{eqnarray}
\end{center} 
where $m_1 =(0,0,0,0)$, $m_2=(0,0,0,1/2)$ and $m_3=(0,0,1/2,0)$.
The sum in $\chi_{12}$ extends over the fifteen Gorel 
quadruple\cite{igusas}
a sequence of four even characteristics which form a syzygous sequence.
Next consider the hyperelliptic curve for a genus 2g+2 surface in the
form
\begin{equation}
y^2= \Pi_{i=1}^{2g+2}(x-e_i)= P_{2g+2}(x, e_i),\;\;e_i \neq e_j,\;
for\;i \neq j,
\label{hyp1}
\end{equation}
respectively in the genus case as
\begin{equation}
y^2= \Pi_{i=1}^{6} (x-e_i) = P_6(x, e_i),\;\;e_i
\neq e_j,\;for\;i \neq j,
\label{hyp2}
\end{equation}
which represents the double cover of the sphere branched over 2g+2
points,
respectively 6 points.
Note that $\theta_m = D^{1/8}$ , where D is the discriminant of $P_{6}$.
On the curve (\ref{hyp1}) one can attach a lattice generated by the 
pair $(I,\Omega)$,
where  I is a $g \times g$ identity matrix and $\Omega$ is the period 
matrix\cite{far}.
The surface (\ref{hyp1}) equipped with the pair $(I, \Omega)$ is 
called a Jacobian. 
We denote by $(a_1, \cdot, a_g,b_1,\dots,b_g)$ the canonical homology
basis on (\ref{hyp1}). In addition, we can define by 
$(\zeta_1, \dots,  \zeta_g)$ as the dual homology basis. Note that 
the dual
homology basis can be defined in terms of $\zeta_i =x_i dz/y,
\;i=1,\cdots, g$. 
In fact by taking the map from the Jacobian to the complex numbers
we define the $\cal \theta$ functions 
\begin{equation}
{\cal \theta}\left( \begin{array}{c}a\\b\end{array}\right)
= \sum_{n \in Z^{g+1}}exp[(n+a)^T \Omega(n+a) + 2 \pi i(n+a) \cdot 
(z+b)]
\label{klio12}
\end{equation}
with the usual build in $Sp(2g,Z)$ invariance. 
One might now start to feel what we are trying to prove. We will prove
that under certain conditions {\em all N=1 (0,2) orbifold
compactifications of the heterotic string can be described by   
certain genus two curves, the binary sextics.}. 
In particular once the moduli coming from a 
heterotic string vacum are known, we
can describe the precise form of the Riemann surface that they line
on. 
\newline
However something is still missing from our discussion.
The element that we require is that there is a birational  
correspondence \cite{siegel} between the projective varieties 
associated with the 
graded ring of even projective invariants of binary sextics 
and with the graded ring of modular forms.
In simple terms that means that the projective variety
linked with the graded ring of even projective invariants
of binary sextics is a compactification of moduli of curves 
of genus two.

So, by keeping in mind the analogy with the branched sextic 
curve coming from 
double covering of ${\P}^2$ in the previous sections, we conjecture the
following theorem :

{\bf Theorem} {\em The projective variety associated with the even 
projective invariants 
of binary sextics, for 
a particular (0,2) $N=1$ heterotic string theory vacuum,
represents the Riemann surface with manifest Sp(4,Z) invariance 
that the moduli live.}
Lets us explain in more detail this issue.
The projective invariants A, B, C, D, of the binary sextics
have degrees 2,4,6,10. So if we symbolize by $\varphi_1, \varphi_2,
\dots,\varphi_6$ the six roots of the sextic 
$s_oX^6+s_1X^5+\dots +s_6$ and we denote their difference 
$\varphi_i -\varphi_j$ by (ij) the invariants take the following form
\begin{eqnarray}
A(s) =s_o^2 \sum_{fifteen}(12)^2 (34)^2 (56)^2\nonumber\\
B(s) =s_o^4 \sum_{ten} (12)^2(23)^2(31)^2(45)^2(56)^2(64)^2\nonumber\\
C(s) =s_o^6 \sum(12)^2(23)^2(31)^2(45)^2(56)^2(64)^2(14)^2
(25)^2(36)^2\nonumber\\
D(s) =s_o^{10} \Pi_{j<k} (jk)^2.
\label{askler1}
\end{eqnarray}
Because any sextic can be brought in the general form
\begin{equation}
X(X-1)(X-\lambda_1)(X-\lambda_2)(X-\lambda_3)
\label{asdf321}
\end{equation}
we can replace each of the three lambdas in (\ref{asdf321})
by some theta functions of zero argument, namely
\begin{equation}
\lambda_1= (\frac{\theta_{1100} \theta_{1000}}{\theta_{0100} 
\theta_{0000}})^2,\; \lambda_2=(\frac{\theta_{1001} \theta_{1100}}{
\theta_{0001} \theta_{0100}})^2,\;
\lambda_3=( \frac{\theta_{1001} \theta_{1000}}{\theta_{0001} 
\theta_{0000}})^2 ,
\end{equation}
where 
\begin{equation}
\theta_{g_1 g_2 h_1 h_2}\left(\begin{array}{cc}\tau_1&\epsilon\\
\epsilon &\tau_2\end{array}\right)=
\sum_{n=0}^{\infty} \frac{2^{2n}}{(2n)!}\frac{d^n}{d\tau_1^n}
\theta_{g_1 h_1} (\tau_1) \frac{d^n}{d\tau_2^n} \theta_{g_2 h_2}(\tau_2)
\epsilon^{2n}
\label{aas1}
\end{equation}
theta functions of genus two.
It can be proved that $\lambda_1, \lambda_2, \lambda_3$ can be expanded
in terms of even powers of $\epsilon$ when 
$\epsilon$ is small.
 As a result all the variables
in eqn. (\ref{asdf321}) are fixed and its roots may be calculated.   
Moreover the invariants A, B, C, D are expressed in terms 
of $\epsilon$ and $\lambda$'s. It is remarkable that the following
relations hold
\begin{equation}  
D/A^5 \propto \epsilon^{12}, (B/A^2)^3 \propto j(\tau_1) j(\tau_2)
\epsilon^{12}, ((3C-AB))/A^3)^2\propto
(j(\tau_1)-j(\tau_2))\epsilon^{12}.
\label{asda1}
\end{equation}
or in precise form
\begin{equation}
I_4=B,\;\; I_6=\frac{1}{2}(AB-3C),\;\; I_{10}=D,\;\; 
I_{12}=AD,\;\; I_{35}=5^3 D^2E.
\label{asda}
\end{equation}
That means that the 
uniformation parameters
 of the sextic (\ref{asdf321}) are fixed 
genus two elements therefore possessing
manifest $Sp(4,Z)$ invariance and therefore confirming our theorem.
In its non-perturbative form the equation for the sextic $P_6$ 
involves the parametric relation (\ref{asda}).  
Note that from eqn. (\ref{asda1}) one can notice that the 
fundamental invariant of the full theory are expressed in terms 
of products of j-invariants. 
What we have not discuss is the expansion of genus two theta functions 
(\ref{aas1}) in terms of the $\epsilon$ parameter, our Wilson lines,
represents exactly the fact that
the space of projective varieties corresponding to the invariants A, B,
C, D has been blown up \cite{siegel} such that the Jacobian variety of 
the genus two
curve has degenerate to products of elliptic curves.
The blow up process is necessary since the projective variety
(\ref{hyp2}), does not include apriori the Siegel 
fundamental domain. 

The correspondence with the heterotic string
comes after identifying $\tau_1=T$, $\tau_2=U$, $\epsilon=Wilson\;line$.  
At the points that the discriminant of the projective variery
of (\ref{hyp1}) degenerates 
both T, U and the Wilson line A, B are involved in a non-trivial relation.
 That means that at the
point
where the discriminant of eqn. (\ref{asdf321}) vanishes one or more of the
moduli
may be fixed therefore breaking non-trivially space-time supersymmetry.

Alternatively thinking, eqn. (\ref{hyp2}) supplied with the projective
invariants A, B, C, D at the limit of small $\epsilon$,
represents the $(0,2)$ perturbative expansion of 
the $N=2$ sector of the 
4$\cal D$ 
heterotic string vacuum.

Lets us now make a comment regarding the 8-dimensional
compactifications of F-theory. 
By taking the equation
\begin{equation}
(branched\;sextic)=0
\label{opili1234}
\end{equation}
we may describe the Riemann surface responsible for the 
8-dimensional $K_3$ fibrations as double cover onto $\P^2$.
Those compactifications that are characterized by the 
vanishing locus have embedded $Sp(4,Z)$ invariance
and may represent how 7-branes degenerate.     
Of course the real test of our conjecture at this point is to 
test this result in its four dimensional F-theory counterparts involving
compactifications on a three-fold. 
This task will be performed in a future work. 

\section{One loop 4D heterotic prepotential for subgroups of $PSL(2,Z)$}

$N=2$ heterotic string theories in four dimensions come from 
compactification of the ten dimensional heterotic string 
on the $K_3 \times T^2$. 
In the simplest case the effective action of light modes 
is invariant under the 
classical modular group is $SL(2,Z)_T \times SL(2,Z)_U$.
In this case the formulation of the six dimensional F-theory compactification, 
prior to further compactification of a $T^2$ torus,  has been given
in \cite{mova1,mova2}. Let us consider the usual F-theory/heterotic duality map
in eight dimensions\cite{fmowi}.
When compactifying the F-theory side on a $K_3$ surface $S_F$ admitting 
an elliptic fibration this 
becomes dual at the heterotic string compactified on a $T^2$ torus.
At the degeneration limit the j-invariant of the F-theory elliptic curve
becomes identical to the j-invariant of the heterotic elliptic curve.      
If we further compactify on an another $K_3$ surface both dual sides 
then what effectively happens is that the statement about monodromy
of the F-theory elliptic curve translates into a statement about
targe space duality. So by the use of the Mordell-Weyl group action
on the F-theory elliptic curve in the original eight dimensional
compactification of F-theory we can control the target space duality
group
in four dimensions. 

Let us now consider that our F-theory elliptic curve contains points
that are associated to a cyclic subgroup of order 2 that a generator has
not been 
chosen. In this case the F-theory has monodromy $\Gamma_o(2)$ and the    
4${\cal D}$ $N=2$ theory coming from
compactification of the heterotic string compactified on the 
$K_3 \times T^2$
is invariant under the target space 
duality group ${\Gamma_o(2)}_T \times {\Gamma_o(2)}_U$.
The third derivative of the prepotential of the $N=2$ vector
multiplet theory is given by
\begin{equation}
f_{TTT}= \frac{96 i}{\pi} \frac{(\Phi_T (T))^2}{\Phi (T)(\Phi(T) -
\Phi(\frac{i}{\sqrt{2}}))} \frac{\Phi^{\frac{1}{2}} (U)}{\Phi_U (U)}
\frac{ {(\Phi(U)- \Phi(\frac{i}{\sqrt{2}}))}^{\frac{1}{2}} }{ 
(\Phi(U) -\Phi(T))}.
\label{prepoo1}
\end{equation}
At the limit that T drifts towards $U_g=\frac{aU +b}{cU + d}$, where g 
an $\Gamma_o(2)$ element
\begin{equation}
f_{TTT} \rightarrow \frac{-2i}{\pi} \frac{1}{T-U_g} (cU + d)^2,
\label{prepoo2}
\end{equation}
only when
\begin{equation}
F(U_g) >> F(\frac{i}{\sqrt{2}}).
\label{prepoo3}
\end{equation}
At this limit the one loop K\"ahler metric exhibits the usual
logarithmic singularity
\begin{equation}
G_{T{\bar T}}^{(1)} \rightarrow \frac{2}{\pi} \ln|T -U_g|^2 
G_{T {\bar T}}^{(0)}\;\;.
\label{prepoo3}
\end{equation}
When our theory is invariant under the group ${\Gamma^o(2)}_T 
\times {\Gamma^o(2)}_U$, $F_{TTT}$ becomes
\begin{equation}
f_{TTT}= \frac{96 i}{\pi} \frac{(\phi_T (T))^2}{\phi (T)(\phi(T) -
\phi({i\sqrt{2}}))} \frac{\phi^{\frac{1}{2}} (U)}{\phi_U (U)}
\frac{ {(\phi(U)- \phi({i \sqrt{2}}))}^{\frac{1}{2}} }{ 
(\phi(U) -\phi(T))},
\label{prepoo4}
\end{equation}
where
\begin{equation}
\phi = \frac{\Delta(z/2)}{\Delta(z)}  
\label{prepoo5}
\end{equation}
the Hauptmodul for $\Gamma^o(2)$.

{\bf Note Added}

The results of this work have been presented in the author's talk at
SUSY '98 and its transparencies are available at 
{\bf http://hepnts1.rl.ac.uk/SUSY98/}.
We note that exactly the same day appeared in the hep-th archive the 
work of \cite{moo}
where singular $K_3$ surfaces are used in the description of the 
attractor mechanism.
\newline
{\bf Acknowledgements}

We would like to thank 
D. Zagier for usuful discussions and B. Mazur for a suggestion.
Also we would like to thank the Isaak Newton Institute for
 Mathematical 
Sciences at Cambridge for its hospitality and the kind use of its 
facilities
in the context of my  
Isaac Newton Junior Membership and the Newton Institute Workshop
on Computational Results of Arithmetic Geometry.
We are especially grateful to U. Persson for sending us a copy of
 \cite{ena}
and for important discussions.

\end{document}